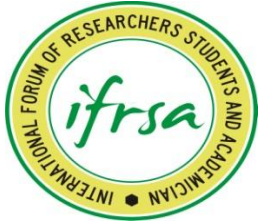

# IIJDWM

**Journal homepage: www.ifrsa.org**

# Web Usage Mining: Pattern Discovery and Forecasting

Dhanamma Jagli[1], Sangeeta Oswal[2]
[1,2]Assistant Professor, Department of MCA,
V.E.S. Institute Of Technology, Mumbai-74, India.

**ABSTRACT**

Web usage mining: automatic discovery of patterns in clickstreams and associated data collected or generated as a result of user interactions with one or more Web sites. This paper describes web usage mining for our college log files to analyze the behavioral patterns and profiles of users interacting with a Web site. The discovered patterns are represented as clusters that are frequently accessed by groups of visitors with common interests. In this paper, the visitors and hits were forecasted to predict the further access statistics.

**Keywords:** Web usage mining, Web log, Cluster, Forecast

## 1. INTRODUCTION

The web revolution had a profound impact on the way we search and find information at home and work place. The web has also become an enormously important tool for communicating ideas, conducting business and entertainment. The exponential growth in the web has been unbelievable. The web is estimated to grow to almost billions of pages. Web as become the number one source for information for internet users. Web usage mining is the third category in web mining. This type of web mining allows the collection of Web access information for Web pages. This usage data provides the paths leading to accessed Web pages. This information is often gathered automatically into access logs via the Web server. This category is important to the overall use of data mining for companies and their internet/ intranet based applications and information access.

In this paper, we describe about sample log file of our college web server. web log is collected and analyzed to form clusters on the basis of hits, visitor and bandwidth in order forecast future web access pattern(hits) so that access statics like bandwidth etc., could be improved to accommodate new visitors.

## (I) WEB USAGE MINING(WUM)

Web mining is the application of data mining techniques to find interesting and potentially useful knowledge from web data it is normally expected that either the hyperlink structure of the web or the web log data or both have been used in the mining process. Web mining mainly can be divided into three categories s as follows

- *Web content Mining*
- *Web Structure Mining*
- *Web usage Mining*

### a) Web Usage Mining

It deals with understanding user behavior in interacting with the website and it tries to make sense of the data generated by the web surfer's sessions and behaviors'. Analyzes access patterns of user to improve response .the mined data often includes data log of users interactions with the web sites. The logs include the web server logs, proxy server logs and browser logs. The logs include information about the referring pages, user identification, time user spend, and sequence of pages visited. Information is also collected via cookies files.

### b) WUM Phases

There are three main tasks for performing WUM— preprocessing, pattern discovery and pattern analysis [1]. These are briefly explained as follows.

**Preprocessing:** Commonly used as a preliminary data mining practice, data preprocessing transforms the data into a format that will be more easily and effectively processed for the purpose of the user. The different types of pre processing in WUM are— usage, content, and structure preprocessing.

**Pattern Discovery:** WUM can be used to uncover patterns in server logs but is often carried out only on samples of data. The mining process will be ineffective if the samples are not a good representation of the larger body of data. The various pattern discovery methods are— Statistical Analysis, Association Rules,

Clustering, Classification, Sequential Patterns, and Dependency Modeling.

**Pattern Analysis:** The need behind pattern analysis is to filter out uninteresting rules or patterns from the set found in the pattern discovery phase. The most common form of pattern analysis consists of a knowledge query mechanism such as SQL. Content and structure information can be used to filter out patterns containing pages of a certain usage type, content type, or pages that match a certain hyperlink structure. Mechanism used: SQL, OLAP, visualization etc

c) **Web log structure –Example**

151.48.123.70
- -
[08/Dec/2007:00:00:41 -0800]
"GET /order/?ref=002 HTTP/1.1"
 200
3467
"www.smsync.com"
"Mozilla/4.0 (compatible; MSIE 7.0; Windows NT 5.1
Web log field
IP 151.48.123.70

Name: the name of the remote user (usually omitted or replaced by dasd"-")
Login: login of the remote user (usually omitted or replaced by dasd"-")
Date/Time/TZ: [08/Dec/2007:00:00:41 -0800]
Request, status code, object size, referrer, and user agent

d) **Types of log used in web usage Mining is:**

*Transfer /Access log:* Contains detailed information about each request that the server receives from user's web browsers.
*Agent log:* Lists the browser that people are using to connect to server.
*Referrer log:* Contained the URL of pages on other sites that link to your pages that is if a user gets to one of the server pages by clicking on link from another site, URL of the site will appear in this log.
*Error log:* keeps a record of error and failed requests.

**(II) OVERVIEW OF WUM**

In Web Usage Mining, knowledge is extracted and used to predict and forecast based on web log file as shown in below Fig1.

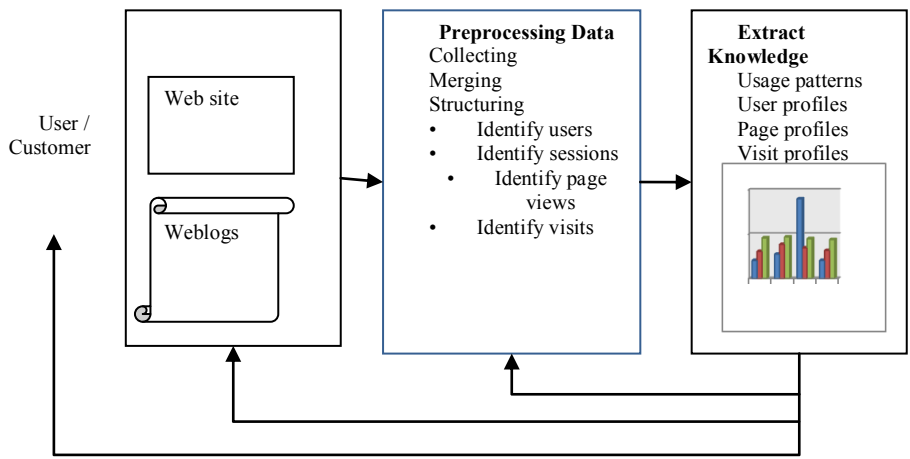

**Figure 1: Extracting knowledge from Web usage Data**

In our proposed work,, we found the pattern in the log file access to improve the web site performance .the knowledge extracting workflow comprises of preprocessing the data to identify useful information like sessions, user ,visit ,page view, bandwidth etc.This information can be mined using various data mining techniques like clustering , Association Rule Mining , Forecasting etc .

In this paper, clustering is applied to group on the basis of hits, visitor and bandwidth which is explained in detail below. The forecasted values used to predict the usage patterns. The results were proved experimentally shown below.

**2. RESULT**

In this paper log file from college web server was collected and analyzed using web usage mining tool: Web Log Expert .The output generated from this tool was further given to SPSS to form clusters and forecast predicated value for bandwidth and visitors as shown in below Fig2.

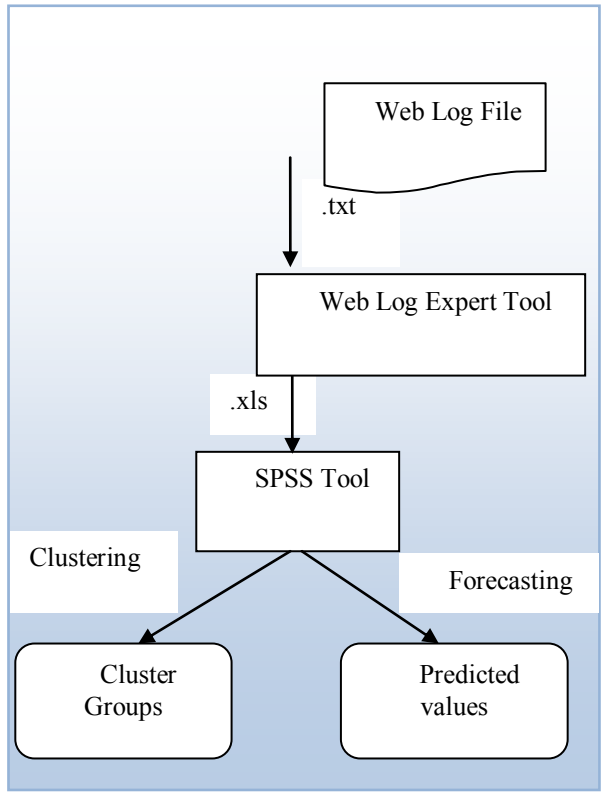

**Figure 2: Design Steps**

*Step 1:* web log expert analyses the log file to produce web usage details that includes hits, average visitor per day ,total page view, total bandwidth etc

**Hits**
Total Hits 39,986
Visitor Hits 39,986
Average Hits per Day 6,664
Average Hits per Visitor 166.61
Cached Requests 0
Failed Requests 0
**Page Views**
Total Page Views 21,346
Average Page Views per Day 3,557
Average Page Views per Visitor 88.94
**Visitors**
Total Visitors 240
Average Visitors per Day 40
Total Unique IPs 149
**Bandwidth**
Total Bandwidth 610.52 MB
Visitor Bandwidth 610.52 MB
Spider Bandwidth 0 B
Average Bandwidth per Day 101.75 MB
Average Bandwidth per Hit 15.63 KB
Average Bandwidth per Visitor 2.54 MB
*Step 2:* The output which comprises of pages viewed hits, incomplete request, visitors and bandwidth is given to SPSS.Using hierarchical clustering, 3 clusters were formed based on hits, visitors, bandwidth using Euclidian distance and centroid linkage. The snapshot of cluster membership is shown below

| Cluster Membership | |
|---|---|
| Case | 3 Clusters |
| 1:http:/ /www.google.com/ | 1 |
| 2:http:/ /www.google.co.in/compressiontest/ gzip.html | 1 |
| 3:http:/ /go.microsoft.com/fwlink/ ?LinkId=74005 | 1 |
| 4:http:/ /runonce.msn.com/runonce3.aspx | 1 |
| 5:http:/ /ocsp.verisign.com/ | 1 |
| 6:http:/ /ocsp.thawte.com/ | 1 |
| 7:http:/ /www.rediff.com/ | 2 |
| 8:http:/ /www.gmail.com/ | 1 |
| 9:http:/ /in.msn.com/?rd=1 | 2 |

*Cluster 1:* It is for max no. of hits, visitors but bandwidth is comparatively less than the other two clusters
*Cluster 2:* It is for the site which has less visitors and hits than cluster 1 but bandwidth is more
*Cluster3:* It is for the site with highest bandwidth but less hits and visitors
*Step 3*:At the same time using ARIMA model ,the predicated values for hits and visitors is forecasted for future access statics of the site .The sample output for forecasted values as shown in the below Fig.3.

| Host | Hits | Visitors | Bandwidth | Year | Month | Date | Pre_hits | Pre_visitor |
|---|---|---|---|---|---|---|---|---|
| http://www.google.com/ | 90 | 66 | 94 | 2011 | 1 | Jan-11 | 2624 | 36 |
| http://www.google.co.in/compressiontest/ gzip.h | 67 | 52 | 53 | 2011 | 2 | Feb-11 | 405 | 66 |
| http://go.microsoft.com/fwlink/ ?LinkId=74005 | 59 | 40 | 27 | 2011 | 3 | Mar-11 | 384 | 52 |
| http://runonce.msn.com/runonce3.aspx | 59 | 40 | 842 | 2011 | 4 | Apr-11 | 377 | 40 |
| http://ocsp.verisign.com/ | 34 | 28 | 67 | 2011 | 5 | May-11 | 377 | 40 |
| http://ocsp.thawte.com/ | 102 | 27 | 141 | 2011 | 6 | Jun-11 | 355 | 28 |
| http://www.rediff.com/ | 53 | 26 | 1501 | 2011 | 7 | Jul-11 | 415 | 27 |
| http://www.gmail.com/ | 28 | 26 | 15 | 2011 | 8 | Aug-11 | 372 | 26 |
| http://in.msn.com/?rd=1 | 36 | 25 | 2496 | 2011 | 9 | Sep-11 | 350 | 26 |
| http://mail.yahoo.com/ | 28 | 24 | 17 | 2011 | 10 | Oct-11 | 357 | 25 |
| http://google.com/pagead/drt/ui/ | 80 | 23 | 56 | 2011 | 11 | Nov-11 | 350 | 24 |

Figure 3:Sample Output

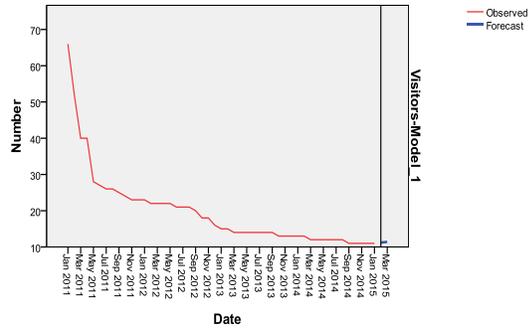

Figure 4: Graph for forecasted results

### 3. CONCLUSIONS

Web usage mining has emerged as the essential tool for realizing more personalized user-friendly and business-optimal Web services the proposed methods were successfully tested on the log files for cluster formation and forecasting the values for visitor and hits. The results which were obtained after the analysis were satisfactory and contained valuable information about the log files. Analysis of above obtained information proved WUM as a powerful technique in Website management and improvement. However, the data obtain from web server log may not deal with issues like caching and proxy servers.